# Template assisted self-assembly of individual and clusters of magnetic nanoparticles

*Giovanni A. Badini Confalonieri*[1,*], *Victor Vega*[2], *Astrid Ebbing*[1], *Durgamadhab Mishra*[1], *Philipp Szary*[1], *Victor M. Prida*[2], *Oleg Petracic*[1], and *Hartmut Zabel*[1]

1- Experimentalphysik IV – Festkörperphysik, Ruhr-Universität Bochum, 44801 Bochum, Germany.

2- Univ. Oviedo, Dept Phys, Oviedo 33007, Spain.

AUTHOR EMAIL ADDRESS: giovanni.badini@rub.de

**ABSTRACT** The deliberate control over the spatial arrangement of nanostructures is the desired goal for many applications as e.g. in data storage, plasmonics or sensor arrays. Here we present a novel method to assist the self-assembly process of magnetic nanoparticles. The method makes use of nanostructured aluminum templates obtained after anodization of aluminum disks and the subsequent growth and removal of the newly formed alumina layer, resulting in a regular honeycomb type array of hexagonally shaped valleys. The iron oxide nanoparticles, 20 nm in diameter, are spin coated onto the nanostructured templates. Depending on the size, each hexagon site can host up to 30 nanoparticles. These nanoparticles form clusters of different arrangements within the valleys, such as collars, chains, and hexagonally closed islands. Ultimately, it is possible to isolate individual nanoparticles. The strengths of magnetic interaction between particles in a cluster is probed using the memory effect known from the coupled state in superspin glass systems.

KEYWORDS Iron Oxide nanoparticles, self-assembly, aluminum template, magnetic superspin glass.

MANUSCRIPT TEXT

Magnetic nanoparticles (NPs) are recognized as promising systems with a high potential for several technological applications, e.g. in magnetic data storage media and spintronic devices or in plasmonic, photonic or nanomedical systems[1,2,3,4]. Presently, large efforts are being devoted in order to investigate and control their chemical and physical properties[5]. The magnetic properties of NP assemblies are proven to depend on their degree of packing and ordering, which might result in different particle to particle interactions having both short range and long range effects[6,7,8,9]. Various synthesis methods have been developed in order to control the formation, packing and self-assembly of NPs at the nanoscale, ranging from inorganic materials (as e.g. metallic or metal-oxide NPs), macromolecules, and biomolecules on 1D and 2D surfaces, and on 3D patterned substrates[10,11,12,13,14].

Many of these methods, including lithography-based techniques and/or other complex nano-engineering routes, require very sophisticated experimental equipment and high operation costs when accurate sub-100 nm ordered structures are to be achieved[15]. However, one low-cost, easy to manipulate and well-reproducible technique to form highly ordered 2D to 3D patterned nanostructures over large areas is based on nanoporous anodic alumina membranes (NAAMs)[16,17].

In this work NAAMs are utilized as a new scheme for the self-assembly of NPs that successfully allows to control the relative distance between isolated clusters of magnetic nanoparticles as well as of individual NPs. The new composite nanomaterial is obtained by combining two nanostructured materials, namely iron oxide NPs, prepared by a chemical route,[18] and nanostructured aluminum templates obtained through an electrochemical route[19,20].

Patterned templates of nanostructured alumina were obtained by following an electrochemical procedure. A schematic drawing of all synthesis steps is shown in Figure 1. High purity Al foils (Goodfellow, 99.999%) were cleaned by sonication in isopropanol and ethanol for 5 minutes. The surface of the starting Al substrate was then smoothed by double side electropolishing in a 1:3 vol. perchloric acid and ethanol mixture at 5ºC during 15 minutes, and by applying a constant dc potential of 20 V between the sample and a Pt mesh. The electropolished Al foils were mounted in a homemade electrochemical anodization cell serving as the anodic electrode. A Pt mesh was kept at a distance of about 1cm from the anode and was used as the cathode. Anodization was performed under potentiostatic conditions using different applied voltages and acidic electrolytes that were kept at fixed temperatures and moderately stirred during the entire process.

This procedure allows us to prepare several self-ordered NAAMs having different pattern parameters. In all cases the final nanostructured Al templates consist of a highly ordered honeycomb lattice, the center of each hexagon defining the pore site. The anodization process allows to vary the pore size and the pore-pore distance of this close packed structure. Three sets of NAAMs were prepared, acting as precursors of the patterned Al substrates. Two of them were prepared following the well-established Mild-Anodization (MA) process[19,20], while another was prepared through the Hard-Anodization (HA) process[21]. In the latter case the use of high anodization potentiostatic voltages provides higher nanoporous alumina growth rates. The experimental conditions for the synthesis procedure of NAAMs templates are listed in Table 1.

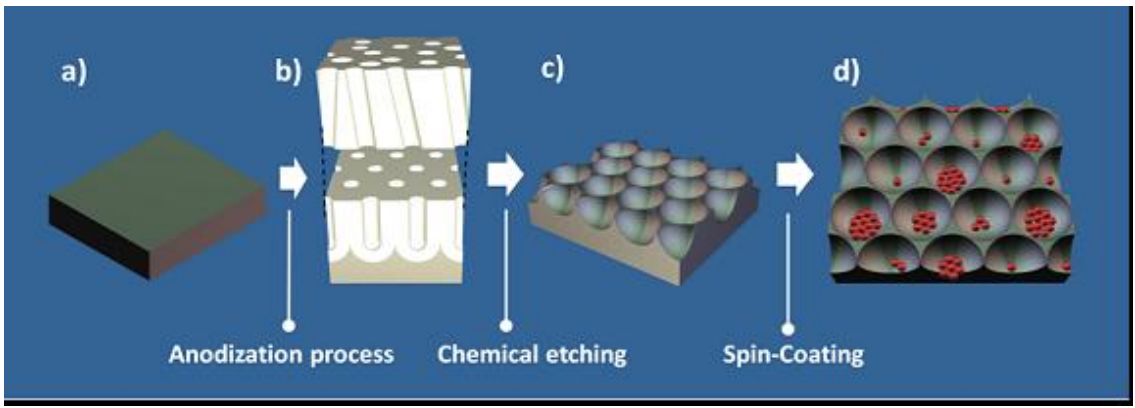

**Figure 1**: Schematic drawing of the different stages followed during the synthesis procedure of the nanocomposite: a) High purity electropolished Al foil. b) Nanoporous alumina membrane. c) Highly ordered Al nanotemplate. d) Nanocomposite formed by individuals and clusters of nanoparticles self-assembled into the nanocups of the highly ordered Al patterned template.

In all cases, during the controlled "bottom-up" anodization process of the NAAMs, the initial randomly disordered pores growing on the alumina surface develop a highly ordered hexagonal structure via self-assembly of the nanopores, which penetrate all the way through the alumina layer (schematically shown in Fig. 1b)[22]. In order to achieve a high degree of order of both the nanoporous alumina layer (Fig. 1b) and the aluminum templates (Fig. 1c), the anodization time was extended up to 1 h and 96 h for HA NAAMs and MA NAAMs templates, respectively. After stopping the anodization process, the resulting layer of the nanoporous alumina membrane was removed by selective chemical etching in an aqueous solution of $CrO_3$ 0.18 M and $H_3PO_4$ (0.61 M), to obtain the final nanostructured Al patterned substrate (Fig. 1c), which is used as a template for the magnetic NP assembly. The exposed top-surface of the nanostructured Al template forms a honeycomb lattice of semispherical valleys, as shown schematically in Fig. 1c, with sizes ranging from 50 nm up to 250 nm depending on the parameters of the anodization process. Nanostructured templates with a disk diameter of 17 mm have been produced.

Commercial iron oxide NPs were purchased from NN-Labs. They were prepared by thermal decomposition of metallic oleates and have a nominal diameter of 20 nm with a size distribution of about 10%. The as received NPs were annealed at 170° C in air for 20 min in order to obtain predominantly single phase maghemite ($\gamma$-$Fe_2O_3$), as reported in[23]. The NPs were dissolved in toluene and spin coated on top of the Al patterned substrates at 300 rpm for 3 sec and subsequently at 2000 rpm for 30 sec, following the method described in [23]. A monolayer film of the same kind of NPs, spin coated on a (100) Si substrate, was also prepared for comparison.

**Table 1. Synthesis conditions employed during the fabrication process of NAAM templates**

| Process designation | Electrolyte | Anodization V ($V_{dc}$) | Temperature (ºC) |
|---|---|---|---|
| Oxalic M.A. | 0.3 M Oxalic acid | 40 | 1-3 |
| Sulphuric M.A. | 0.3 M Sulphuric acid | 25 | 1 |
| Oxalic H.A. | 0.3 M Oxalic acid | 140 | 0-3 |

Structural characterizations of the samples were performed by means of a FEI Quanta 200 FEG scanning electron microscope (SEM), and the magnetic properties analyzed by zero-field-cooling (ZFC) and field-cooling (FC) magnetization measurements using a Quantum Design MPMS superconducting quantum interference device (SQUID) magnetometer in the temperature range between 330 K and 15 K, and an applied fields up to 50 kOe.

Three different types of nanostructured aluminum templates have been prepared with three different lattice parameters as shown in Figure 2. The nanostructured aluminum is characterized by a honeycomb arrangement of nanocups with diameters d= 280 nm, d= 105 nm and d= 72 nm for samples obtained by Oxalic HA, Oxalic MA and Sulphuric MA, respectively. Obviously the smallest pattern exhibits a number of structural defects.

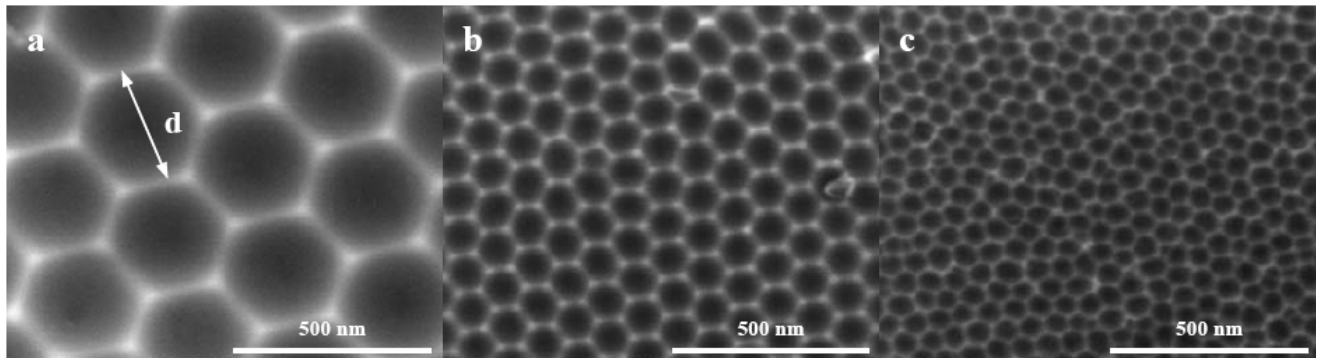

**Figure 2**: SEM images of nanostructured aluminum patterned substrate obtained after anodization in a) Oxalic acid HA, b) Oxalic acid MA , and c) Sulphuric acid MA. Scale shown in all figures corresponds to 500 nm

For simplicity, the composite samples will be labeled as D280, D105 and D72, referring to the size of the nanostructure characteristic of the specific anodization conditions used for their preparation. Spin coating of NPs results in different filling factors depending on the patterned substrate used. SEM observations over the whole surface of the samples, from their center to the outer diameter, showed that for all three templates the packing density within the same sample remains constant over the whole area considered.

After spin coating on the Al templates, the magnetic NPs arrange in small clusters at the bottom of each hexagon valley, as found in the SEM images in Figure 3 b-d. The number of NPs in each hexagon site varies from 2 up to ~30. If the number of particles per cluster is less than 10 particles, then the NPs arrange in different packing geometries, sometimes in closed collars or open chains; for clusters composed of more than 10 NPs a hexagonally ordered packing is observed, similar to the self-assembly of particles spin coated on a polished Si substrate, as shown in Figure 3a. For patterns with the smallest lattice parameter (sample D72) the number of particles per host site varies between one and four, but also a large portion of sites remains empty, as can be found in Fig. 3d.

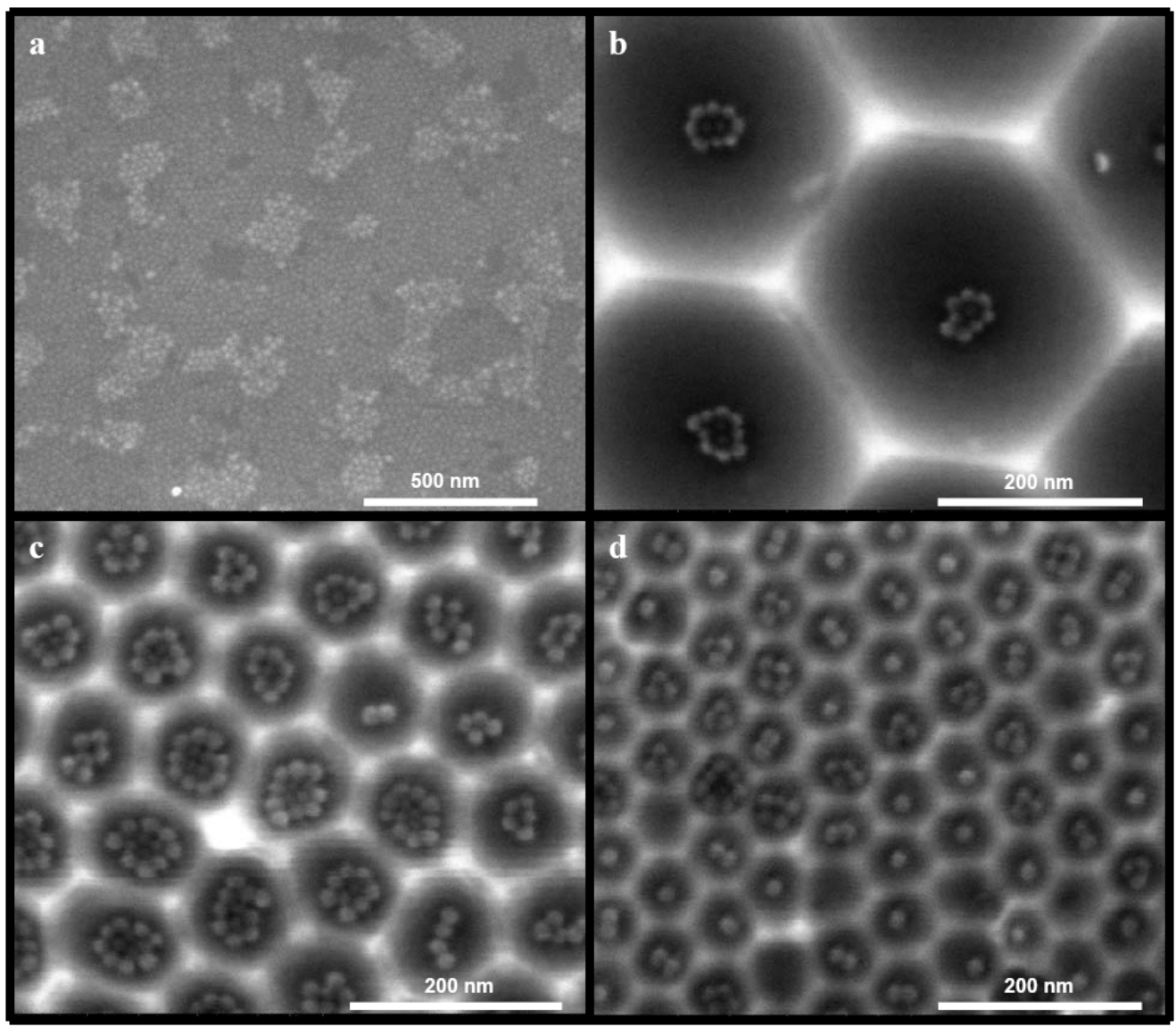

**Figure 3**: SEM images of NP assemblies on top of different patterned Al substrates: a) Oriented (110) Si, b) sample D280, c) sample D105, and d) sample D72

More information can be obtained from the histograms of the NP population on the sites in the three templates shown in Figure 4, where $N_S$ is the number of sites which are occupied by specific clusters, and $N_{NP}$ is the number of NPs per cluster. The histograms were obtained from several images of the samples each covering areas ranging from 1 to 2 μm². The largest distribution of NP-cluster population is found in D280 samples, as can be observed in Figure 4b, where an average of 17 NPs per site is found, with 30% of the occupied sites hosting between 21 and 25 NPs. A narrower size distribution is found in sample D105 (Fig. 4c), where the average number of particles per cluster is 6 and 50% of the occupied sites host between 5 and 8 NPs. Samples D72 have the narrowest size distribution, with an average number of particles per site of 2 and 86% of the occupied sites hosting between 1 and 3 NP, with 46% of the total sites hosting individual NP (Fig. 4c).

Isolated NPs are found to preferentially sit at the centre of each site with an average distance between nearest neighbor nanoparticle of approximately 70 nm, in good agreement with the lattice parameter of the D72 substrate. In this sample, vacant sites account for approximately 11% of the total.

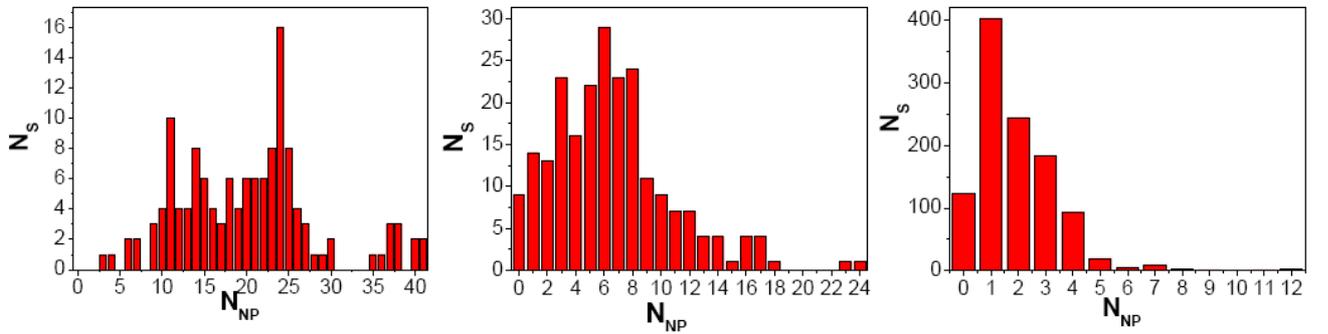

**Figure 4**: NP population histograms for: a) sample D280, b) sample D105, c) sample D72.

The possibility to control the NP-cluster dispersion in an ordered arrangement over the whole surface of a patterned substrate offers a unique scenario for the investigation of competing magnetic interactions in systems with low dimensionalities. In a system of interacting single domain NPs, two main possible magnetic behaviors can be observed.

In the first case, when the interactions between particles are sufficiently small, the magnetic behavior is dominated by the magnetic moments of the individual NPs. In this case, the system behaves superparamagnetic (SPM). [7, 24,25]

In the second case, for a system with a sufficiently high concentration of NPs and consequently stronger particle-particle dipolar interactions, it may be possible to observe a collectively coupled state. Two prominent examples of which are the superspin glass (SSG)[2,26,27,28,29,30] and the superferromagnetic (SFM) state[1,31,32]. The effect of the interparticle dipolar interaction depends on the specific type of arrangement in each cluster as well as on the number of NPs in each cluster.

For the analysis of their magnetic behavior, we have performed field cooling - zero field cooling (FC-ZFC) measurements of the magnetization. For the ZFC protocol the sample was cooled from an initial temperature of $T_i$ = 330 K to a final temperature $T_f$ = 15 K in zero field, then a magnetic field of 50 Oe was applied and the ZFC curve measured upon warming. The FC curve is obtained directly following the ZFC curve upon cooling in the same applied field.

The results are depicted in Figure 5, where for comparison purpose the ZFC-FC magnetization values are normalized to the maximum value of the MZFC curve.

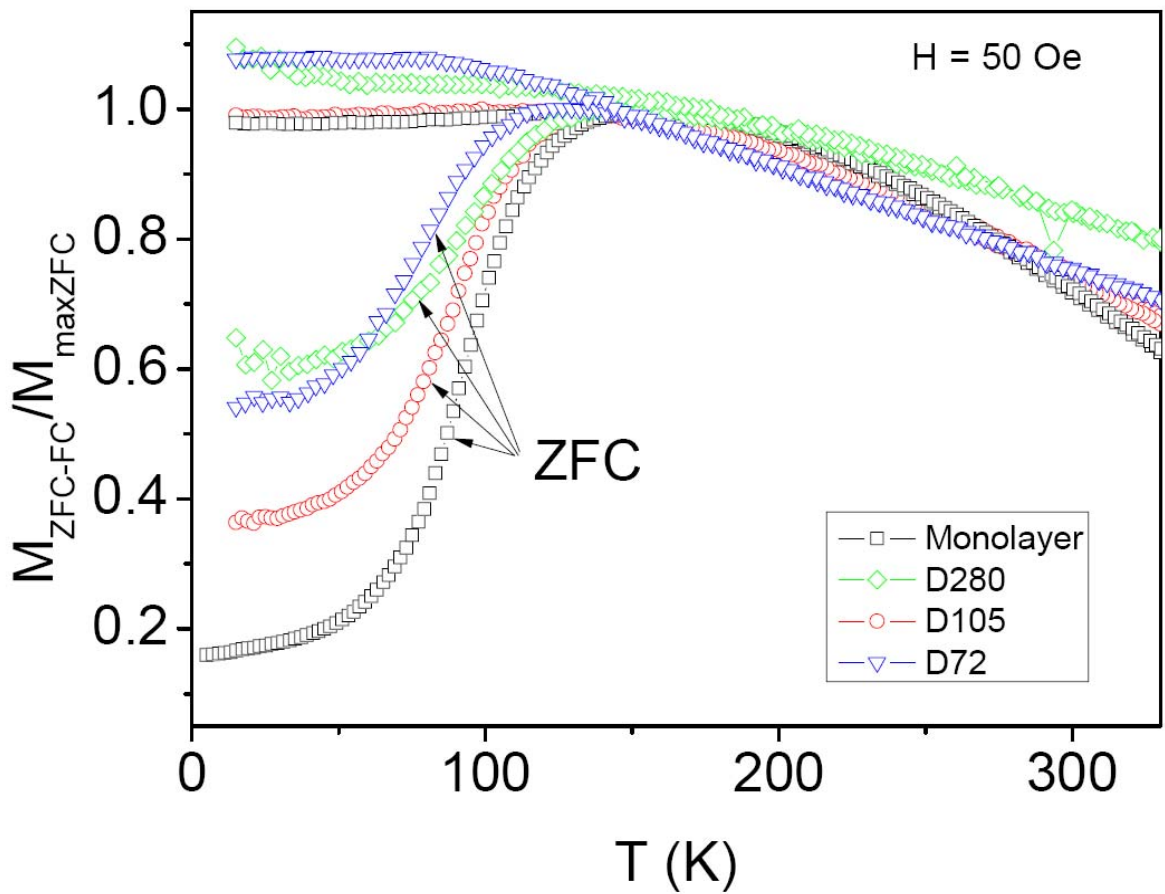

**Figure 5**: Normalized ZFC-FC curves for NPs assembled on: a Si substrate (black squares), sample D280 (green diamonds), sample D105 (red circles), and sample D72 (blue triangle)

The magnetic behavior of the monolayer of γ-$Fe_2O_3$ NPs on the Si-substrate agrees well with the results reported in previous work[23], where a detailed study of the magnetic properties of NP monolayers was performed.

In the present study the shape of the FC magnetization $M_{FC}$ shows features, which are different from those for regular NP monolayers[7,23] that hint to novel behavior. One finds in Fig. 5 that with decreasing the temperature below the blocking temperature $T_b$ the magnetization for the clusters of NPs is either constant or drops slightly. This feature has been reported to originate from dipolar interactions between NPs such as is the case of SSG systems[2,9,33,34,35]. The $M_{FC}$ is also observed to decrease, when the particles are arranged in clusters, as in the case of sample D105. This can be explained as follows: although the clusters themselves are too small to display SSG behavior, the average edge to edge distance between clusters (~50 nm) is of the same order of magnitude as the average cluster size. Therefore, dipolar cluster-to-cluster interactions can be expected. However, the lower slope of the $M_{FC}$ curve in sample D105 suggests a decreasing strength of the dipolar interactions compared with the NP monolayer on a Si substrate.

On the other hand, when NP-clusters are placed further apart, as in the case of sample D280, the long range interaction becomes negligible and $M_{FC}$ follows the usual superparamagnetic behavior of increasing $M_{FC}$ with decreasing temperature below $T_b$.

Sample D72 presents some unique features. In first place $M_{FC}$ remains constant in the temperature range from $T_b$ to $T_f$. This can be attributed to a coexistence of both non-interacting individual NPs as well as dipolarly coupled clusters of NPs separated with an average edge to edge distance of ~40 nm. A further difference is found in the value of $T_b$ = 125 K, which suggests that $T_b$ is decreased, when individual particles are successfully isolated. In fact, in the case of the NP monolayer, sample D280, and D105, $T_b$ remains constant at approximately 150 K, indicating that the blocking temperature does not depend on the long range SSG-like interactions. It should be mentioned that even within isolated small clusters of particles, dipolar interactions between the constituent particles are expected. This would explain the different $T_b$ values between samples D280 and D72.

A unique and characteristic feature of SSG (and more generally spin glass) systems is the memory effect of the ZFC curve[27,29,36]. In this type of measurement, the sample is first cooled from $T_i$ to $T_f$ in zero applied external field. The magnetization is then recorded as a function of temperature in the interval $T_f$ to $T_i$ at an applied field of $H$ = 50 Oe as a regular ZFC curve. In a second step, the measurement is repeated with the difference that during cooling in zero field, the sample is hold at a stop temperature $T_s$ = 90 K, with $T_f < T_s < T_i$, for a sufficiently long waiting time $t_w$ = 10000 s. The two ZFC curves (conventional $M_{ZFC}$ and $M_{ZFC}(T_s)$ after halting at $T_s$) are then compared. In SSG systems, the sample will "remember" the treatment received at $T_s$ showing a deviation in the $M_{ZFC}(T_s)$ curve with respect to the regular ZFC curve.

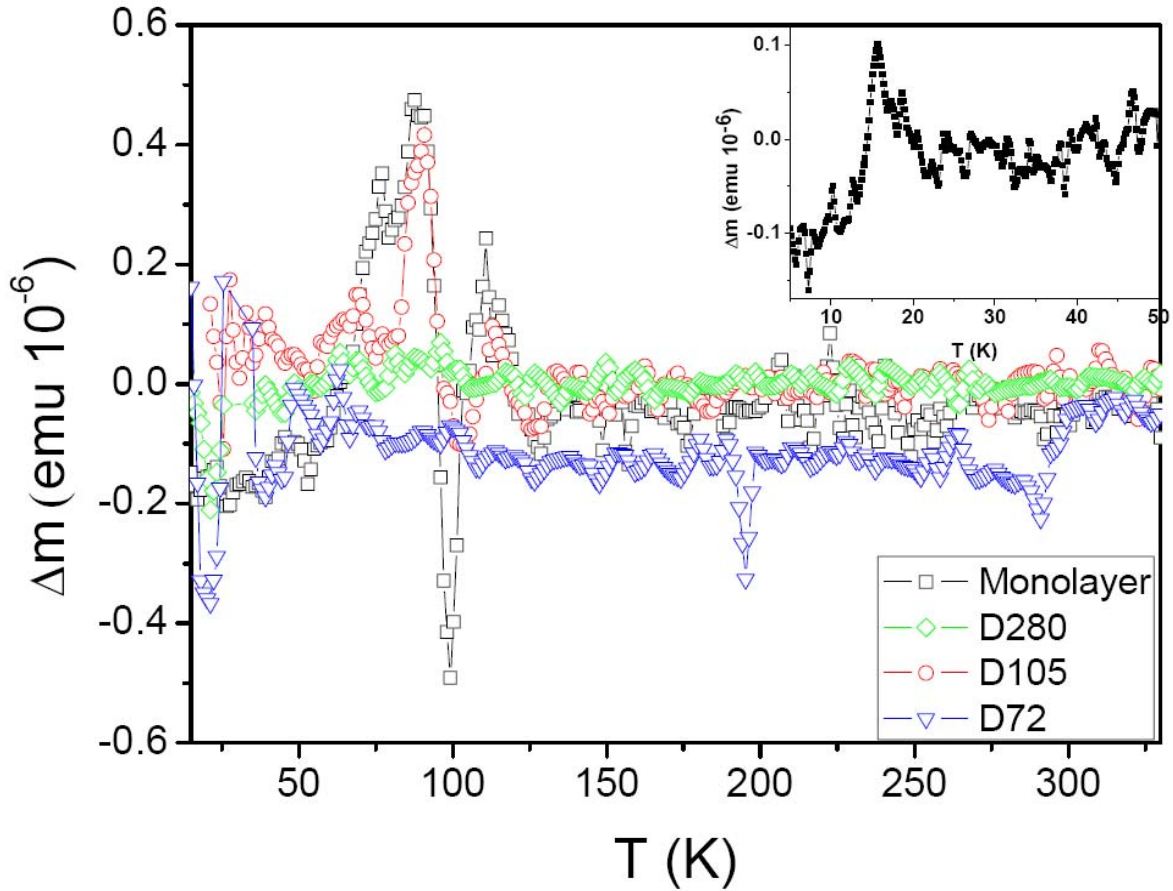

**Figure 6**: Δm difference between two ZFC curves in the memory effect experiment for NPs assembled on: a Si substrate (black squares), sample D280 (green diamonds), sample D105 (red circles), and sample D72 (blue triangles). The stop temperature was in all cases $T_s$ = 90 K. For comparison purpose

the memory effect observed in a SSG system of Co NPs is shown in the inset. Here the stop temperature was $T_s = 20$ K

Measurements of the memory effect were performed on the three template cluster samples as well as on the NP monolayer sample. Figure 6 shows the difference $\Delta m = m_{ZFC} - m_{ZFC}(T_s)$. It can be found that both the NP monolayer film and sample D105 display a memory effect near T = 90 K, confirming a SSG behavior of the two systems.

On the other hand, the samples D280 and D72 show no hint of any memory effect. Consequently no collective SSG behavior is observed. This is in agreement with the results obtained from the analysis of the ZFC-FC curves shown in Figure 5, where long range dipolar interactions were only observed for the monolayer system and sample D105. It should be noted that the $\Delta m$ memory curves of the monolayer and sample D105 show an unexpected feature of a downward pointing peak near $T_s$. Usually, the memory effect in SSG systems is characterized by a relatively broad positive peak in $\Delta m$[34,37,38].

To show that this feature is not simply an artifact of our measurement, we have studied another system of Co NPs for comparison. The Co NPs are formed by ion beam sputtering of Co on a $Al_2O_3$ buffer layer similar to the technique used in[29,39,40]. Details on the preparation and behavior of these samples will be published elsewhere. This type of system is known to display SSG behavior.[29,30]

Memory ZFC measurements were performed on this type of Co NPs with $T_i = 50$K, $T_f = 5$K, $T_s = 20$K, $t_w = 10000$ s and $H = 20$ Oe. The results are shown in the inset of Figure 5. One finds the usual positive peak near to the stopping temperature.

Therefore we tentatively assume that the appearance of a downward peak at Ts is a unique feature of the Fe-oxide NP systems. The origin of this feature is still unclear and further work is ongoing to clarify this point.

The results obtained in this work prove that nanostructured aluminum disks with a honeycomb pattern of different lattice parameters can be successfully used as templates for the self-assembly of Fe-oxide

NPs spin-coated on the surface. By choosing the appropriate template lattice parameters the NPs arrange, with a high ordering degree, into small clusters or can even be successfully isolated on individual host sites. Modifying the distance among clusters and/or individual NPs can destroy the long-range magnetostatic coupling and hence inhibit the formation of a collective SSG state otherwise often found in monolayer films of close-packed magnetic NPs. The preparation method proposed in this work can easily be extended to a different variety of NP systems, offering new possibilities for the study of both fundamental properties of magnetic NPs as well as technological applications.

Acknowledgements: This work was supported by the Ruhr-Universität Bochum through the Research Department IS$^3$/HTM. Financial grants under FEDER, Spanish MICINN and FICYT research projects Nos: MAT2009-13108-C02-01, FC09-IB09-131 and MAT2010-20798-C05-04 are recognized. The scientific support of SCTs from University of Oviedo, particularly to the Nanoporous Membranes Unit, is also acknowledged.

REFERENCES

(1) Naito, K.; Hieda, H.; Sakurai, M.; Kamata, Y.; Asakawa, K.; *IEEE Trans. Magn.* **2002**, 38, 1949

(2) Terris, B. D.; Thomson, T.; *J. Phys. D: Appl. Phys.* **2005,** 38, R199

(3) Black, C. T.; Murray, C. B.; Sandstrom, R. L.; Sun, S.; *Science* **2000**, 290, 1131

(4) Reiss, G.; Hutten, A.; *Nature Materials* **2005**, 4, 725

(5) Sun, S. H.; *Adv. Mater.* **2006**, 18, 393

(6) Puntes, V. J.; Gorostiza, P.; Aruguete, D. M.; Bastus, N. G.; Alivisatos, A. P.; *Nature Mater.* **2004**, 3, 263

(7) Dormann, J. L.; Fiorani, D.; Tronc, E.; *Advances in Chemical Phys.* **1997**, 98, 283

(8) Sun, S. H.; Murray, C. B.; *J . Appl. Phys.* **1999**, 85, 4325

(9) Jonsson, T.; Mattsson, J.; Djurberg, C.; Khan, F. A.; Nordblad, P.; Svedlindh, P.; *Phys. Rev. Lett.* **1995,** 75, 4138.

(10) Gao, H.; Gosvami, N. N.; Deng, J.; Tan, L. S., Sanders, M.; *Langmuir* **2006**, 22, 8078

(11) Cheng, J. Y.; Zhang, F.; Chuang, V. P.; Mayes, A. M.; Ross, C. A.; *Nano Letters* **2006**, 6, 2099

(12) Pan, B.; Cui, C.; Ozkan, C.; Xu, P.; Huang, T.; Li, Q.; Chen, H.; Liu, F.; Gao, F.; He, R.; *J. Phys. Chem.* **2007**, 111, 12572

(13) Toster, J.; Iyer, K. S.; Burtovyy, R.; Burgess, S. S. O.; Luzinov, I.A.; Raston, C. L.; *J. Am. Chem. Soc.* **2009**, 131, 8356

(14) Grzelczak, M.; Vermant, J.; Furst, E. M.; Liz-Marzán, L. M.; *ACS Nano* **2010**, 4, 3591

(15) Cho, J.H.; Gracias, D.H.; *Nano Letters* **2009**, 9, 4049

(16) Ko, H.; Chang, S.; Tsukruk, V.V.; *ACS Nano* **2009**, 3, 181

(17) Liu, K.; Nogues, J.; Leighton, C.; Masuda, H.; Nishio, K.; Roshchin, I. V.; Schuller, I. K.; *Appl. Phys. Lett.* **2002**, 81, 4434

(18) Park, J.; An, K.; Hwang, Y.; Park, J. G.; Noh, H. J.; Kim, J. Y.; Park, J. H.; Hwang N. M.; Hyeon, T.; *Nature Mater.* **2004**, 3, 891

(19) Masuda, H.; Fukuda, K.; *Science* **1995**, 268, 1466

(20) Vazquez, M.; Pirota, K. R.; Navas, D.; Asenjo, A.; Hernandez-Velez, M.; Prieto, P.; Sanz, J. M.; *J. Magn. Magn. Mater.* **2008**, 320, 1978

(21) Lee, W.; Ji, R.; Gösele, U.; Nielsch, K.; *Nature Mater.* **2006**, 5, 741

(22) Prida, V.,M.; Pirota, K. R.; Navas, D.; Asenjo, A.; Hernandez-Velez, M.; Vazquez, M.; *J. Nanoscience and Nanotechnology* **2007**, 7, 272

(23) Benitez, M. J.; Mishra, D.; Szary, P.; Feyen, M.; Lu, A. H.; Agudo, L.; Eggeler, G.; Petracic, O.; Zabel, H.; *J. Phys.: Cond. Mat.* **2010** submitted, ArXiv:1010.0938

(24) Néel, L.; *Ann. Geophys.* **1949**, 5, 99

(25) Brown, .W. F.; *Phys. Rev.* **1863**, 130, 1677

(26) Jonsson, P.E.; *Adv. Chem. Phys.* **2004**, 128, 191.

(27) Mamiya, H.; Nakatani, I.; Furubayashi, T.; *Phys. Rev. Lett.* **1999**, 82, 4332.

(28) Sahoo, S.; Petracic, O.; Kleemann, W.; Nordblad, P.; Cardoso, S.; Freitas, P.P.; *Phys. Rev. B* **2003**, 67, 214422.

(29) Petracic, O.; *Superlatt. Microstr.* **2010**, 47, 569

(30) Batlle, X.; Labarta, A.; *J. Phys. D* **2005**, 35, R15.

(31) Chen, X.; Sichelschmidt, O.; Kleemann, W.; Petracic, O.; Binek, C.; Sousa, J.B.; Cardoso, S.; Freitas, P. P.; *Phys. Rev. Lett.* **2002**, 89, 137203.

(32) Hauschild, J.; Elmers, H. J.; Gradmann, U.; *Phys. Rev. B* **1998**, 57, R677.

(33) Petracic, O.; Chen, X.; Bedanta, S.; Kleemann, W.; Sahoo, S.; Cardoso, S.; Freitas, P.P.; *J. Magn. Magn. Mater.* **2006**, 300, 192

(34) Parker, D.; Dupuis, V.; Ladieu, F.; Bouchaud, J.-P.; Dubois, E.; Perzynski, R.; Vincent. E.; *Phys. Rev. B* **2008**, 77, 104428

(35) Mamiya, H.; Nakatani, I.; Furubayashi, T.; Phys. Rev. Lett. **1998**, 80, 177

(36) Mydosh, J.; *Spin Glasses: An Experimental Introduction* **1983** Taylor and Francis, London

(37) Sasaki, M.; Jönsson, P. E.; Takayama, H.; Mamiya, H.; *Phys. Rev. B* **2005**, 71, 104405

(38) Suzuki, M.; Fullem, S. I.; Suzuki, I. S.; *Phys. Rev B* **2009**, 79, 024418

(39) Sankar, S.; Berkowitz, A. E.; Smith, D. J.; *Appl. Phys. Lett.* **1998**, 73, 535

(40) Kakazei, G. N.; Pogorelov, Y. G.; L. Lopes, A. M.; Sousa, J. B.; Cardoso, S.; Freitas, P. P.; Pereira de Azevedo, M. M.; Snoeck; E.; *J. Appl. Phys.* **2001**, 90, 4044